\documentclass[english,aps,pra,showpacs,eqsecnum,amsmath,amssymb,reprint]{revtex4-1}
\usepackage[T1]{fontenc}
\usepackage[latin9]{inputenc}
\usepackage{babel}
\usepackage{amsmath}
\usepackage{amssymb}
\usepackage{graphicx}
\usepackage{esint}
\usepackage[unicode=true,pdfusetitle,
 bookmarks=true,bookmarksnumbered=false,bookmarksopen=false,
 breaklinks=false,pdfborder={0 0 1},backref=false,colorlinks=false]
 {hyperref}
\usepackage{breakurl}

\makeatletter
 
 \@ifundefined{textcolor}{}
 {%
   \definecolor{BLACK}{gray}{0}
   \definecolor{WHITE}{gray}{1}
   \definecolor{RED}{rgb}{1,0,0}
   \definecolor{GREEN}{rgb}{0,1,0}
   \definecolor{BLUE}{rgb}{0,0,1}
   \definecolor{CYAN}{cmyk}{1,0,0,0}
   \definecolor{MAGENTA}{cmyk}{0,1,0,0}
   \definecolor{YELLOW}{cmyk}{0,0,1,0}
 }

\usepackage{bm}
\usepackage{amsfonts}
\usepackage{url}
\usepackage{textcase}

\makeatother

\begin{document}

\title{Symmetry breaking and physical properties of the bosonic single-impurity
Anderson model}

\author{\emph{Jesus Herazo Warnes}}

\author{\emph{E. Miranda}}

\affiliation{Instituto de F\'{\i}sica Gleb Wataghin, Rua S\'{e}rgio Buarque de Holanda,
777, CEP 13083-859 Campinas, SP, Brazil}
\begin{abstract}
We show how exact diagonalization of small clusters can be used as
a fast and reliable impurity solver by determining the phase diagram
and physical properties of the bosonic single-impurity Anderson model.
This is specially important for applications which require the
solution of a large number of different single-impurity problems,
such as the bosonic dynamical mean field theory of disordered systems.
In particular, we investigate the connection between spontaneous global
gauge symmetry breaking and the occurrence of Bose-Einstein condensation
(BEC). We show how BEC is accurately signaled by the appearance of
broken symmetry, even when a fairly modest number of states is retained.
The occurrence of symmetry breaking can be detected both by adding
a small conjugate field or, as in generic quantum critical points,
by the divergence of the associated phase susceptibility. Our results
show excellent agreement with the considerably more demanding numerical
renormalization group (NRG) method. We also investigate the mean impurity
occupancy and its fluctuations, identifying an asymmetry in their
critical behavior across the quantum phase transitions between BEC
and `Mott' phases.
\end{abstract}

\pacs{67.85.Bc, 67.85.Jk, 37.10.Gh}

\maketitle

\section{Introduction}

\label{sec:intro}

Strongly correlated impurity models have played an important role
in the field of condensed matter physics. The Anderson \cite{Anderson1961}
and the Kondo \cite{Kondo1964} single-impurity models were at the
heart of the early investigations into the formation of localized
magnetic moments in metals and their effects on thermodynamic and
transport properties of these materials. An extensive arsenal of theoretical
techniques have been developed in an effort to better elucidate these
and related questions \cite{Hewson1993} and a great deal of understanding
has been thereby achieved. More recently, important methods for the
study of \emph{periodic} strongly correlated systems have been developed,
such as the dynamical mean field theory (DMFT) \cite{Georges1996}
and its extensions, which rely heavily on the knowledge base accumulated
in the analysis of the aforementioned impurity models. Indeed, solving
a single-impurity problem is the most challenging part of the DMFT
algorithm.

Since the advent of the possibility of loading extremely cold atoms
onto the effective periodic potential formed by optical lattices \cite{Jaksch1998,JakschZoller2005}
there has been a growing interest in the cross fertilization between
these atomic systems and their conventional condensed matter counterparts.
For one thing, cold atoms in optical lattices are expected to be very
well described by the simplified models used in the condensed matter
context, such as the Hubbard model \cite{JakschZoller2005}. For solids,
by contrast, these models are believed to be at best a bare bones
description which hopefully retains the most important physical features.
Moreover, the parameters in cold-atom optical lattice systems, such
as hopping amplitudes and scattering lengths can often be very flexibly
tuned externally, allowing for the thorough investigation of large
portions of the phase diagrams. Finally, the quantum statistics can
also be switched as bosonic atoms are readily available.

This fruitful interplay between these condensed matter and cold-atom systems has prompted
researchers to attempt to use impurity-model-based approaches as an
analytical tool for cold-atom systems. In particular, the bosonic
version of the dynamical mean field theory (BDMFT) has been developed
\cite{byczukvoll08,SnoekHofstetterChapter,Andersetal2011}. In close
analogy with its fermionic counterpart, this method requires the solution
of a bosonic single-impurity Anderson model (B-SIAM). This model has
been directly studied by the powerfull Wilson numerical renormalization group
(NRG) technique and its phase diagram has been determined \cite{Lee2007,Lee2010}.
Furthermore, applications of BDMFT have also been carried out, using
as impurity solvers exact diagonalization \cite{hubeneretal09,hutong09}
and quantum Monte Carlo \cite{andersetal10,Andersetal2011}. It should
also be mentioned that impurity-like cold-atom set-ups have been proposed
\cite{Recatietal2005} and may also become available (for a realization
in which \emph{different} species occupy the impurity and the bath
regions, see \cite{Zipkes2010}), in which case some version of the
single-impurity model may be directly applicable.

More recently, it has become possible to introduce quench\-ed disorder
into optical lattice systems \cite{horak98,boiron99,diener01,roth03,damski03,gimperlein05,gavish05,massignan06,Fallani2007,Lucioni11},
which may prove useful for the important problem of the interplay
between disorder and interactions \cite{Lee1985a}. The extension
of the BDMFT to the disordered case \cite{other}, however, requires the solution
of a large number of different single-impurity problems for the description
of a single disordered sample, as many as the number of lattice sites
(for a description of the fermionic version, see \cite{Dobrosavljevi'c1998a}).
This makes it all but impossible to use such numerically demanding
single-impurity solvers as the NRG. Therefore, the feasibility of
using the BDMFT to study disordered systems requires the development
and characterization of fast yet reliable single-impurity solvers.

One of the goals of this article is to show that the exact diagonalization
of small clusters is a good solution to this problem, a procedure
which has been successfully applied to the fermionic case \cite{CaffarelKrauth1994,Kajueter1996}.
We will show that it can be efficiently and reliably implemented to
solve the B-SIAM by studying its ground state properties. Moreover,
with an eye towards its application in a BDMFT calculation, we have
thoroughly investigated the appearance of Bose-Einstein condensation
(BEC) in this model as a manifestation of the phenomenon of global
gauge symmetry breaking. Indeed, the B-SIAM is perhaps the simplest
model in which this phenomenon occurs and its reduced Hilbert space
provides a unique setting in which to numerically study it. Although
this is the standard criterion for the detection of BEC, it was not
used in the NRG analysis of the model \cite{Lee2007,Lee2010}. The
same BEC identification which we will show can be done in the B-SIAM
will be useful in the analogous task in a BDMFT calculation. Besides,
the B-SIAM has extended regions in which BEC is present or absent
(the so-called `Mott' phase) even at zero temperature, with quantum
phase transitions between them. We will show how global gauge symmetry
breaking can be used to analyze this quantum phase transition in a
manner analogous to more conventional systems such as magnetic ones,
even though the system sizes used are fairly modest.

We should stress that the question of the best or even the correct
criterion for the occurrence of BEC is not devoid of a certain controversy.
Indeed, while the existence of a macroscopic eigenvalue of the one-particle
density matrix seems to be uncontested as a necessary and sufficient
condition for a BEC \cite{Yang62}, some objections have been raised,
particularly by Leggett \cite{LeggettBook}, as to whether the frequently
used (particularly in the condensed matter literature) criterion of
``spontaneously broken gauge symmetry'' is an adequate alternative.
Much of Leggett's objection seems to be aimed at the \emph{physical
basis} of the infinitesimal symmetry-breaking field usually employed
in this criterion, rather than at the validity of the \emph{mathematical
procedure} it is based upon. Since our main interest here is to show
that this criterion is perfectly adequate for a \emph{numerical} investigation
of the less studied case of an impurity model, we should be quite
safe. Furthermore, it seems clear that, at least in the case of unfragmented
condensates, the criterion of a spontaneously broken gauge symmetry
is both a sufficient and a necessary condition for Bose-Einstein condensation
in bosonic systems (for a specific discussion and review, see \cite{Yukalov2007}).

We will show in this article that, even in fairly small clusters with
a restricted number of bosonic states, a detailed characterization
of the spontaneously broken gauge symmetry of the BEC phase and an
accurate determination of the full phase diagram is possible, which
is in excellent agreement with the much more demanding numerical renormalization
group method. Furthermore, we will analyze how the impurity occupancy
and its fluctuations behave within the phases and through the quantum
phase transitions between them. In particular, we pinpoint a qualitative
difference in the critical behavior of both of these quantities as
one crosses the boundary from BEC to `Mott' as compared to going from
`Mott' to BEC.

The paper is divided as follows. Section \ref{sec:model} is devoted
to the definition of the model and a summary of known results for
the non-interacting as well as the interacting cases. Section \ref{sec:breaking}
expounds on the criterion of global gauge symmetry breaking as a hallmark
of BEC. Details of the numerical procedure are explained in Section
\ref{sec:numerical}. Results on the symmetry breaking occurring in
the model are presented in Section \ref{sub:symmbreakphasediagram},
whereas other local impurity properties are shown in Section \ref{sub:other}.
We draw some final conclusions in Section \ref{sec:Conclusions}.
Some results for the non-interacting limit are relegated to an Appendix.

\section{The bosonic single-impurity Anderson model}

\label{sec:model}

We will focus our attention on the bosonic version of the single-impurity
Anderson model Hamiltonian
\begin{eqnarray}
H & = & \varepsilon_{0}\hat{n}_{0}+\frac{1}{2}U\hat{n}_{0}(\hat{n}_{0}-1)\nonumber \\
 & + & \sum_{k\neq 0}\varepsilon_{k}b_{k}^{\dag}b_{k}^{\phantom{\dagger}}+\sum_{k\neq 0}V_{k}(b_{k}^{\dag}b_{0}^{\phantom{\dagger}}+b_{0}^{\dag}b_{k}^{\phantom{\dagger}}).\label{eq:hamiltonian}
\end{eqnarray}
Here, $b_{k}$ are bosonic annihilation operators for the impurity
($k=0$) and ``bath'' orbitals ($k\neq0$), $\hat{n}_{0}=b_{0}^{\dag}b_{0}^{\phantom{\dagger}}$
is the number operator for the impurity, $\varepsilon_{0}$ is impurity
single-particle energy, and $U$ measures the interaction strength
between bosons inside the impurity orbital. The next two terms of
the Hamiltonian describe, respectively, the bath single-particle orbital
energies $\varepsilon_{k}$ and the hybridization between impurity
and bath states, which occurs with amplitude $V_{k}\in\mathbb{R}$.
We will work in the grand-canonical ensemble at zero temperature and
fixed chemical potential $\mu$. All the single-particle energies
in Equation (\ref{eq:hamiltonian}) are assumed to be measured with respect
to $\mu$. 

Evidently, the physics of the model is strongly dependent on the spectral
properties of the bosonic bath. It is common practice, particularly
in the quantum dissipation literature, to assume a ``soft-gap'' spectral
function for the bath \cite{A.J.Leggett1987}. A power-law dependence
is usually considered \cite{A.J.Leggett1987,Lee2007,Lee2010}, such
that if
\begin{equation}
\Delta\left(\omega-i\delta\right)\equiv\sum_{k}\frac{V_{k}^{2}}{\omega-i\delta-\varepsilon_{k}},\label{eq:bathfunction}
\end{equation}
then
\begin{eqnarray}
\mathrm{Im}\Delta\left(\omega-i\delta\right) & = & \pi\sum_{k}V_{k}^{2}\delta\left(\omega-\varepsilon_{k}\right)\label{eq:spectralfunction}\\
 & = & 2\pi\alpha\omega_{c}^{1-s}\omega^{s}\Theta\left(\omega\right)\Theta\left(\omega_{c}-\omega\right),
\end{eqnarray}
where $\omega_{c}$ is a frequency cutoff, $\alpha$ plays the role
of a dimensionless coupling constant and $\Theta\left(x\right)$ is
the Heaviside step function. The so-called ohmic case corresponds
to $s=1$, whereas $s<1\left(>1\right)$ corresponds to the sub-ohmic
(super-ohmic) regime. 

It is useful to consider first the non-interacting limit, $U=0$.
In this case, the problem can be immediately diagonalized (see the
appendix \ref{sec:non-interacting}). The spectrum $E_{n}$ can be
obtained, e.g., from the impurity site Green's function, and it consists
of the roots of the equation 
\begin{equation}
E_{n}-\varepsilon_{0}=\sum_{k}\frac{V_{k}^{2}}{E_{n}-\varepsilon_{k}}=\Delta\left(E_{n}\right).\label{eq:spectrum}
\end{equation}
In the case of a power-law bath spectral function there is a critical
coupling constant $\alpha_{c}^{0}=s\varepsilon_{0}/2\omega_{c}$ (for
$s>0$), such that there appears a vanishing root $E_{0}=0$ for $\alpha=\alpha_{c}^{0}$
but not for $\alpha<\alpha_{c}^{0}$, as shown in the Appendix. At
this point, the lowest state of the system is macroscopically occupied,
signaling the phenomenon of Bose-Einstein condensation. Further increase
of $\alpha$ renders the system ill-defined (in the grand-canonical
ensemble assumed here), since the lowest state falls below the chemical
potential ($E_{0}<0$).

At finite $U$, the BEC only occurs for $0<s<1$ \cite{Lee2007,Lee2010}.
In this case, for coupling constant values $\alpha<\alpha_{c}\left(\varepsilon_{0},U\right)$,
there is a phase in which the BEC is absent. This phase is adiabatically
connected to its $\alpha=0$ counterpart, in which the impurity is
decoupled from the bath and which is characterized by an integer occupation
of the impurity site. This is very reminiscent of the Mott insulating
phases of the Bose-Hubbard model, hence the name ``Mott phase''. It
should be emphasized, however, that if $\alpha\neq0$ the impurity
occupation deviates from integer values and, in contrast to the Mott
insulating case, it does not exhibit plateaus of constant $\left\langle \hat{n}_{0}\right\rangle $
as a function of $\varepsilon_{0}$ (see Section \ref{sub:other}
below). Therefore, this terminology is used in a loose sense. The
BEC phase is absent for $s>1$ \cite{Lee2007,Lee2010}.

In the NRG study of references \cite{Lee2007,Lee2010}, the transition
from the BEC to the Mott phase was identified from the vanishing of
a gap in the spectrum of low-lying excitations. Indeed, much like
in the $U=0$ limit, the splitting-off of an isolated pole from the
continuum signals the BEC. We will here, however, explore the criterion
of the spontaneous breaking of the (global) gauge symmetry as an alternative
signature of this quantum phase transition, in perfect analogy with
the case of extended bosonic systems.

\section{Spontaneous symmetry breaking}

\label{sec:breaking} 

We now discuss how to look for the BEC phase transition using the
criterion of spontaneously broken gauge symmetry. The original Hamiltonian
(\ref{eq:hamiltonian}) is invariant under the following global gauge
transformation
\begin{equation}
b_{k}\to e^{i\alpha}b_{k},\label{eq:globalgaugetrans}
\end{equation}
which simply reflects the conservation of total particle number $\widehat{N}=\sum_{k}b_{k}^{\dagger}b_{k}^{\phantom{\dagger}}$.
In order to investigate the spontaneous breaking of this symmetry,
one usually introduces a small symmetry breaking field conjugate to
the order parameter. In the BEC case, the latter can be taken to be
$\langle b_{0}\rangle$. We thus modify the Hamiltonian of Equation (\ref{eq:hamiltonian})
as follows ($\varphi\in\mathbb{R}$)
\begin{equation}
H\to H+\varphi\left(b_{0}^{\dagger}+b_{0}^{\phantom{\dagger}}\right).\label{eq:symbreakfield}
\end{equation}
The spontaneous symmetry breaking is signaled by a non-zero value
of the following limit
\begin{equation}
\lim_{\varphi\to0}\lim_{N\to\infty}\frac{1}{N}\vert\langle b_{0}\rangle\vert^{2}\neq0.\label{eq:superfluidp}
\end{equation}
(and it is a necessary and sufficient condition for the existence
of BEC \cite{Yukalov2007}). In Equation (\ref{eq:superfluidp}), $N=\left\langle \widehat{N}\right\rangle $.

This can be illustrated in the somewhat artificial non-interacting
limit. In this case, 
\begin{equation}
H=\varepsilon_{0}b_{0}^{\dagger}b_{0}^{\phantom{\dagger}}+\varphi(b_{0}^{\dagger}+b_{0}^{\phantom{\dagger}})+\sum_{k}\varepsilon_{k}b_{k}^{\dagger}b_{k}^{\phantom{\dagger}}+\sum_{k}V_{k}(b_{k}^{\dagger}b_{0}^{\phantom{\dagger}}+b_{0}^{\dagger}b_{k}^{\phantom{\dagger}}).\label{eq:hamiltonianphi}
\end{equation}
Using the results of the Appendix, it can be shown that the ground
state expectation values are
\begin{eqnarray}
\left\langle b_{0}\right\rangle  & = & \frac{\varphi}{\kappa},\label{eq:b0vphi}\\
\left\langle b_{k}\right\rangle  & = & -\frac{V_{k}}{\varepsilon_{k}}\frac{\varphi}{\kappa},\label{eq:bkvphi}
\end{eqnarray}
where
\begin{equation}
\kappa=\sum_{k}\frac{V_{k}^{2}}{\varepsilon_{k}}-\varepsilon_{0}.\label{eq:kappa}
\end{equation}
We note that the BEC is signaled by the vanishing of the lowest single-particle
energy $E_{0}$ (when measured with respect to the chemical potential).
From Equation (\ref{eq:spectrum}), it is clear that when $E_{0}\to0$,
\begin{equation}
\kappa\to0,\label{eq:kappatozero}
\end{equation}
 and the order parameter $\left\langle b_{0}\right\rangle $ as a
function of $\varphi$ has a diverging slope as a $\varphi\to0$,
such that it tends to a constant in the BEC. Besides, the total number
of bosons is given by (Appendix) 
\begin{equation}
N=\left(1+\sum_{k}\frac{V_{k}^{2}}{\varepsilon_{k}^{2}}\right)\frac{\varphi^{2}}{\kappa^{2}}.\label{eq:Nvphi}
\end{equation}
The limit of Equation (\ref{eq:superfluidp}) in this case is given by
\begin{equation}
\lim_{\varphi\to0}\lim_{N\to\infty}\frac{1}{N}\vert\langle b_{0}\rangle\vert^{2}=\left(1+\sum_{k}\frac{V_{k}^{2}}{\varepsilon_{k}^{2}}\right)^{-1}<1.\label{eq:limb02}
\end{equation}

\section{Numerical method}

\label{sec:numerical}

We now describe the numerical procedure used in the calculations that
follow. The Hamiltonian of Equation (\ref{eq:hamiltonian}) will now be
defined with a finite number $N_{s}$ of bath states
\begin{eqnarray}
H & = & \varepsilon_{0}\hat{n}_{0}+\frac{1}{2}U\hat{n}_{0}(\hat{n}_{0}-1)+\varphi\sum_{n=0}^{N_{s}}\left(b_{n}^{\dagger}+b_{n}^{\phantom{\dagger}}\right)\nonumber \\
 & + & \sum_{n=1}^{N_{s}}\varepsilon_{n}b_{n}^{\dag}b_{n}^{\phantom{\dagger}}+\sum_{n=1}^{N_{s}}V_{n}(b_{n}^{\dag}b_{0}^{\phantom{\dagger}}+b_{0}^{\dag}b_{n}^{\phantom{\dagger}}).\label{eq:hamiltonianfinite}
\end{eqnarray}
Note that we have already included the symmetry breaking field $\varphi$
and it now acts on both impurity and bath states. Since the condensation 
occurs in a single-particle state which is a quantum mixture of all $k$-orbitals
[see Eqs.~(\ref{eq:b0exp}) and (\ref{eq:bkexp}) for the non-interacting case],
it is immaterial whether we couple $\varphi$ to all of them or only to $b_0^{\phantom{\dagger}}$.
Indeed, we have checked
that applying $\varphi$ only to the impurity site does not change
the results that follow in any significant way.

The discretized Hamiltonian (\ref{eq:hamiltonianfinite}) has no conserved
quantities since even the total number of bosons is no longer fixed.
Since the boson spectrum is unlimited the Hilbert space has infinite
dimension. We worked in a cut-off Hilbert space in which there is
a maximum number of bosons

\begin{equation}
\sum_{n=0}^{N_{s}}b_{n}^{\dagger}b_{n}^{\phantom{\dagger}}\le N_{max}.\label{eq:nmax}
\end{equation}
We therefore numerically diagonalized the Hamiltonian \eqref{eq:hamiltonianfinite}
for fixed $N_{s}$ and $N_{max}$ .

The parameters $\{\varepsilon_{n},V_{n}\}$ uniquely determine the
bath spectral function $\mathrm{Im}\Delta(\omega)$, whose support
we will assume is the interval $\left[0,\omega_{c}\right]$, see Equation
(\ref{eq:spectralfunction}). Discretizing the latter set into $N_{s}$
smaller intervals defined by $\left[a_{n-1},a_{n}\right],$ where
$n=1,2,...,N_{s}$, and assuming that $\varepsilon_{n}\in\left[a_{n-1},a_{n}\right]$
it follows that
\begin{eqnarray}
V_{n}^{2} & = & \int_{a_{n-1}}^{a_{n}}\frac{\mathrm{Im}\Delta(x)}{\pi}dx\ \ \left(n=1,2,...,N_{s}\right),\label{eq:vn}\\
V_{n}^{2}\varepsilon_{n} & = & \int_{a_{n-1}}^{a_{n}}\frac{x\mathrm{Im}\Delta(x)}{\pi}dx\ \ \left(n=1,2,...,N_{s}\right).\label{eq:epsilonn}
\end{eqnarray}
Although the use of a purely logarithmic mesh would have been enough,
we have chosen to use a mixed linear-logarithmic set
\begin{equation}
a_{n}=\frac{n}{N_{s}}\frac{\omega_{c}}{\Lambda^{\left(N_{s}-n\right)}},\ \ \ \left(n=0,1,2,\ldots,N_{s}\right).\label{eq:an}
\end{equation}
It is close to the usual logarithmic discretization of the numerical
renormalization group at low energies but does a better job at
describing the high-energy part of the bath spectrum.
The linear discretization is recovered in
the limit $\Lambda\to1$. In the calculations that follow, we have
used $\Lambda=2$.

\section{Results}

We will now show our results for the bosonic single-impurity Anderson
model with a power-law spectral function as in Equation (\ref{eq:spectralfunction}),
with $s=0.4$. This model exhibits two phases: a Bose-Einstein condensed
phase (BEC) and a Mott phase, as shown in \cite{Lee2007,Lee2010}.

\subsection{Symmetry breaking and the phase diagram}

\label{sub:symmbreakphasediagram}

\begin{figure}[h!]
\includegraphics[scale=0.33]{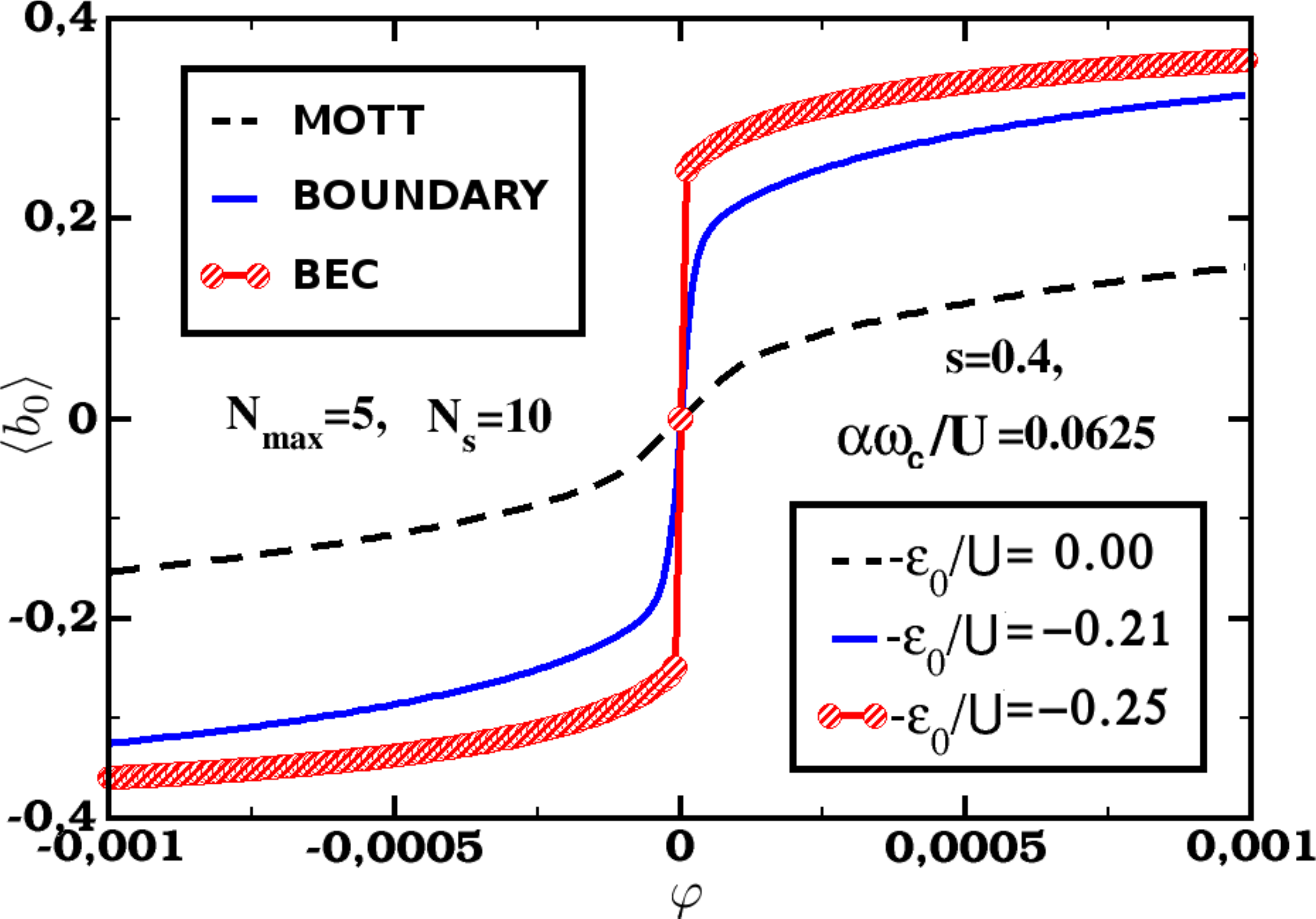} \caption{\label{fig:bversusphi} The superfluid order parameter
$\left\langle b_{0}\right\rangle $ as a function of the symmetry
breaking field $\varphi$ for three different values of $\varepsilon_{0}/U$:
the dashed black line $\left(\varepsilon_{0}/U=0\right)$ corresponds
to the Mott phase, the red circles $\left(\varepsilon_{0}/U=-0.25\right)$
to the BEC phase and the full blue line $\left(\varepsilon_{0}/U=-0.21\right)$
to the boundary between the two phases. The slope at $\varphi=0$
is finite within the Mott phase and infinite at the phase boundary.
We have used $N_{s}=10$, $N_{max}=5$, $U=0.5\omega_{c}$, $\alpha\omega_{c}/U=0.0625$
and $s=0.4$.}
\end{figure}

We have computed the value of the order parameter $\left\langle b_{0}\right\rangle $
as a function of the symmetry breaking field $\varphi$ for a coupling
to the bosonic bath of $\alpha\omega_{c}/U=0.0625$ and interaction
strength $U=0.5\omega_{c}$. According to the NRG results \cite{Lee2007,Lee2010},
for these values of $\alpha$ and $U$ the system may find itself
in either the Mott or the BEC phases, depending of the value of $\varepsilon_{0}$.
As shown in Figure \ref{fig:bversusphi}, for $\varepsilon_{0}/U=0$,
$\left\langle b_{0}\right\rangle $ smoothly extrapolates to zero,
with a finite slope, as $\varphi\to0$. This is characteristic of
a non-condensed phase with no broken global gauge symmetry, the Mott
phase of the model. At $\varepsilon_{0}/U=-0.21$, however, even though
$\left\langle b_{0}\right\rangle $ still vanishes in this limit,
it does so with a very large slope, effectively infinite within our
numerical accuracy. This signals the boundary with the BEC phase and
points to a second order phase transition, consistent with the NRG
results \cite{Lee2007,Lee2010}. Inside the BEC phase ($\varepsilon_{0}/U=-0.25$),
the superfluid parameter has a step discontinuity (again within our
numerical accuracy) across the $\varphi=0$ line. The similarity with
the behavior of a ferromagnet in a uniform external field is striking,
highlighting the common underlying symmetry breaking mechanism in
both cases.

\begin{figure}[h!]
\includegraphics[scale=0.4]{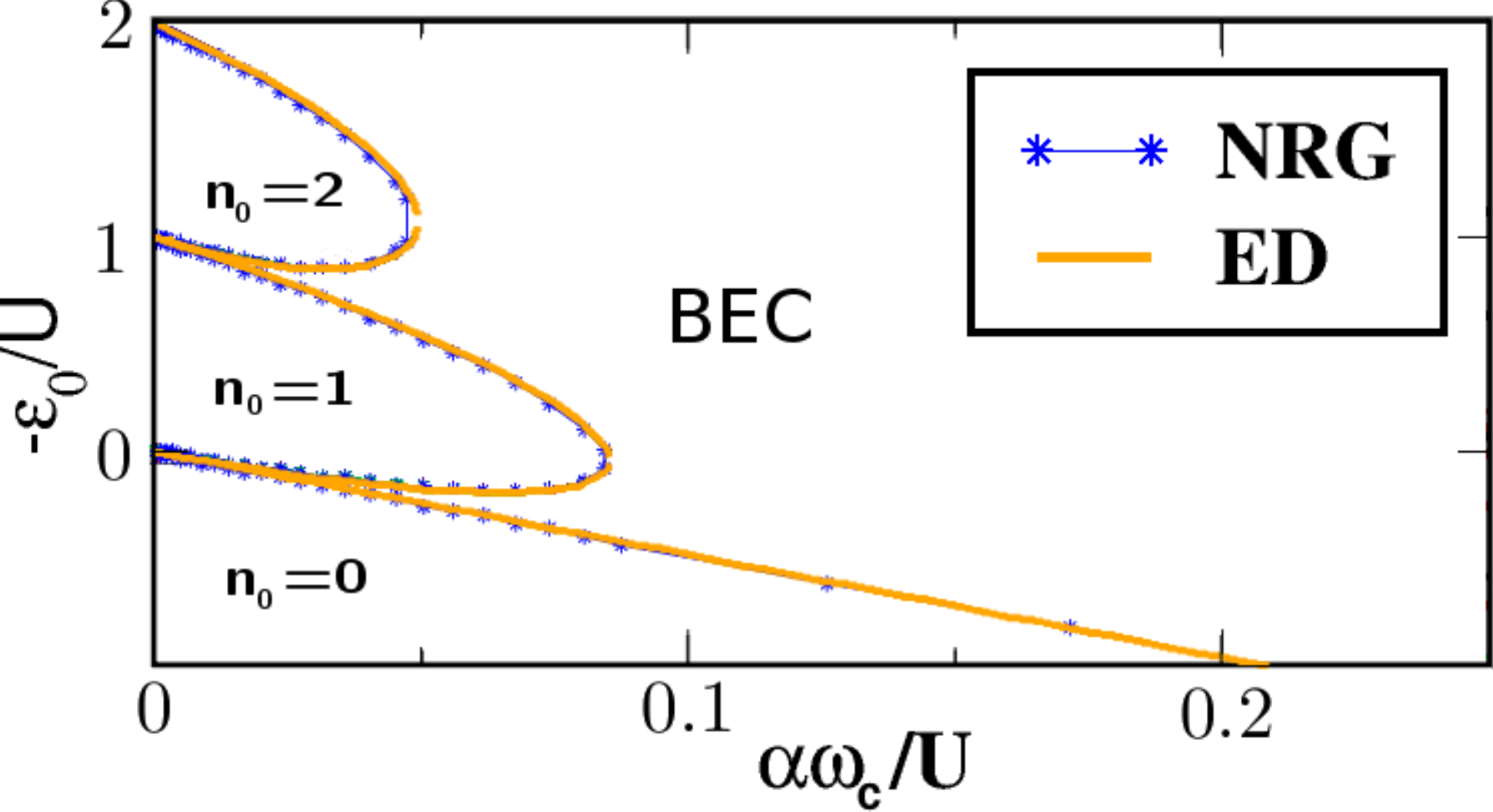} \caption{\label{fig:phasediagram} Phase diagram of the bosonic
single-impurity Anderson model for $s=0.4$ and $U/\omega_{c}=0.5$
(other parameters as in Figure \ref{fig:bversusphi}). The blue line
with symbols is the phase diagram calculated with the NRG method \cite{Lee2007,Lee2010}
and the full orange line was determined using the criterion of global
gauge symmetry breaking and exact diagonalization. See text for more
details.}
\end{figure}

By looking for the points of infinite slope of the $\left\langle b_{0}\right\rangle $
versus $\varphi$ curves one can then map out the phase diagram of
the model. We have done so in the $\varepsilon_{0}$ versus $\alpha$
plane for $U=0.5\omega_{c}$. In practice, we have set the phase boundary as the point
at which the slope reaches $10^4$. 
The same phase diagram had been previously
obtained with the much more powerful NRG method using the `gap closure'
criterion (see Section \ref{sec:model}) \cite{Lee2007,Lee2010}.
The NRG parameters used were $\Lambda=2$ (for a purely logarithmic
discretization), 10 to 20 bosonic states for each added site/iteration
and 100 to 200 states kept from each iteration to the next. In our
exact diagonalization method, we have used $N_{s}=10-20$ and $N_{max}=2-6$.
In Figure \ref{fig:phasediagram}, a comparison between the results
obtained with the two methods is shown. Despite the simplicity of
the present procedure and small number of states retained, 
the agreement is remarkable. We have also verified that the susceptibility
criterion we used tracks very closely the opening of the gap discussed
in Section \ref{sec:model}. The assignment
of average occupations $n_{0}=\left\langle \hat{n}_{0}\right\rangle $
to the phases, as shown in Figure \ref{fig:phasediagram}, will be discussed
later in Sec. \ref{sub:other}.

Another interesting quantity is the phase susceptibility, 
\begin{equation}
\chi_{\varphi}=\lim_{\varphi\to0}\dfrac{\partial\vert\langle b_{0}\rangle\vert}{\partial\varphi}.\label{eq:phasesuscept}
\end{equation}
As discussed in Section \ref{sec:breaking}, this quantity is finite
in the absence of BEC and diverges at the critical point at which
BEC first appears, see Equations \eqref{eq:b0vphi}, \eqref{eq:kappa}
and \eqref{eq:kappatozero}. The analogous quantity in the case of
a ferromagnet is the magnetic susceptibility, which is also finite
in the paramagnetic phase and diverges at the (second-order) phase
transition. Thus, we expect $\chi_{\varphi}$ to be finite in the
Mott phase and to diverge at the Mott-BEC boundary. That the transition
is indeed second-order was confirmed in Refs. \cite{Lee2007,Lee2010}.
This behavior is apparently consistent with the gross features of Figure
\ref{fig:bversusphi}, but it is important to determine how it is
affected by our Hilbert space truncation.

\begin{figure}[h!]
\includegraphics[scale=0.35]{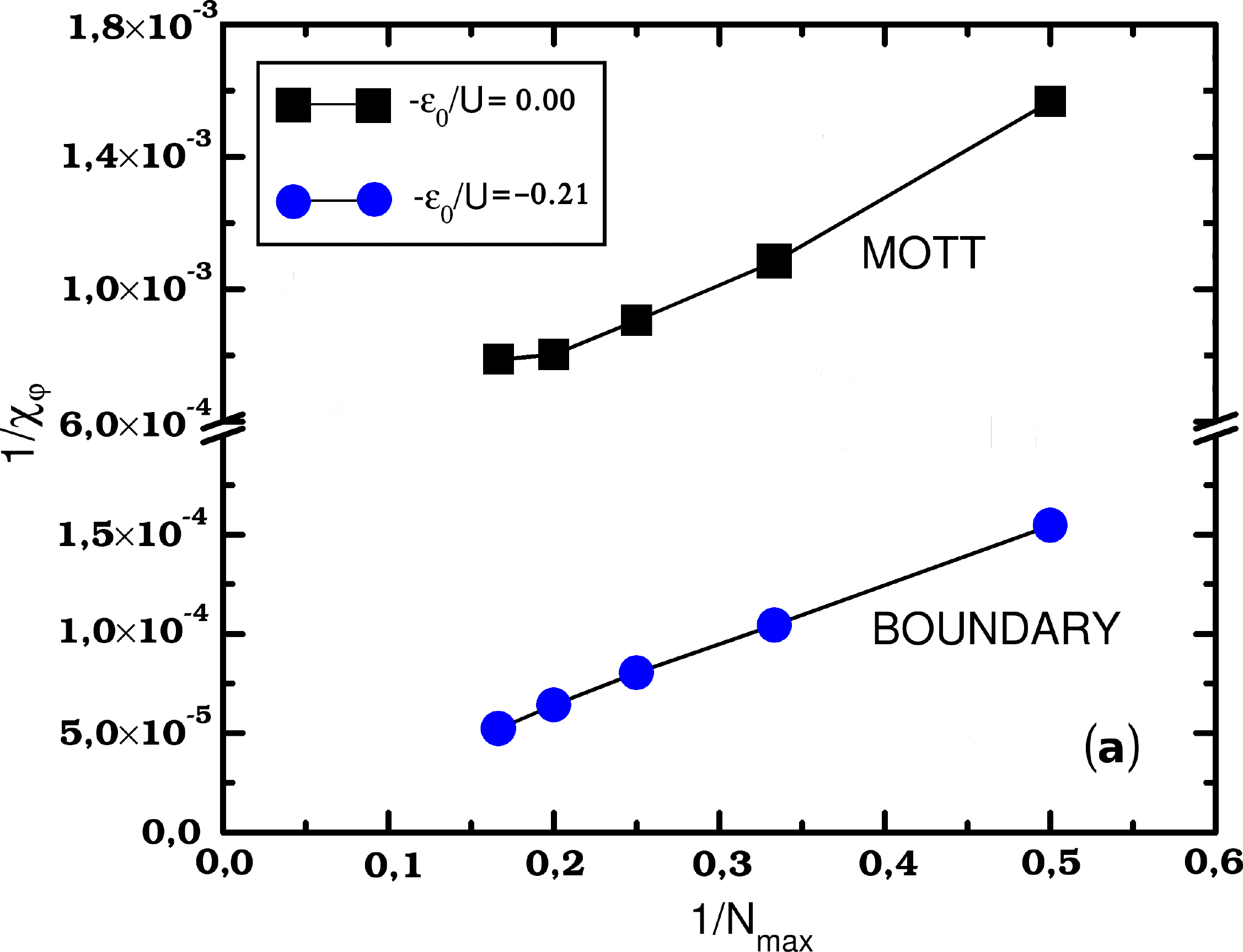}

\includegraphics[scale=0.33]{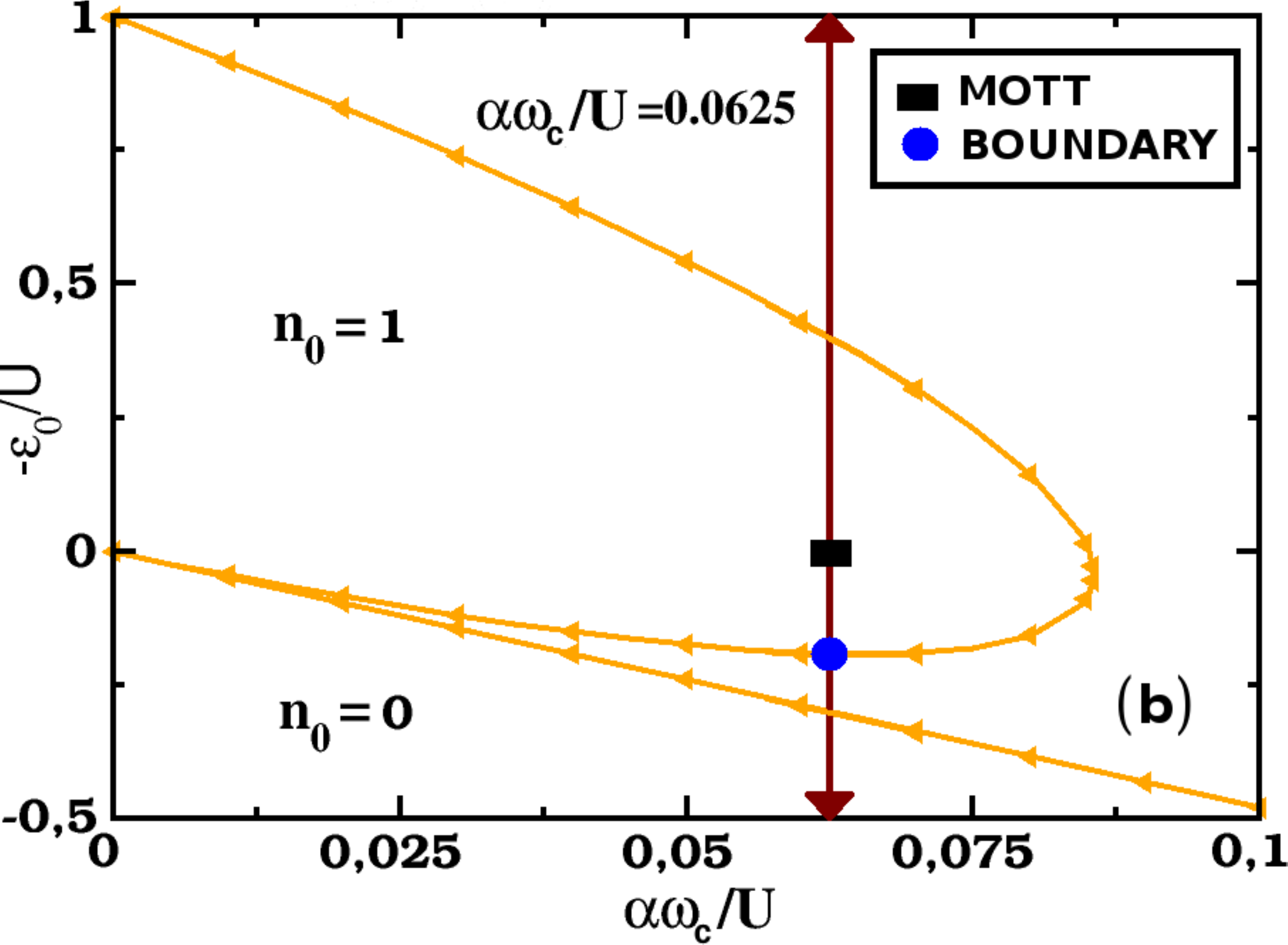}

\centering{}\caption{\label{fig:susceptibility} (a) Phase susceptibility
as a function of $1/N_{max}$ for two values of $\varepsilon_{0}$:
for $\varepsilon_{0}/U=0$, the system is in the Mott phase and $\chi_{\varphi}$
extrapolates to a constant when $N_{max}\to\infty$; the value $\varepsilon_{0}/U=0.21$
marks the boundary between Mott and BEC phases and correspondingly
gives to a divergent $\chi_{\varphi}$ in the limit of infinite $N_{max}$.
Other parameters are the same as in Figure \ref{fig:bversusphi}. (b)
Position in the phase diagram of the parameters in (a).}
\end{figure}

In Figure \ref{fig:susceptibility}(a), we show the inverse phase susceptibility
as a function of $1/N_{max}$. It can be seen that $\chi_{\varphi}$
extrapolates as expected in the limit of $N_{max}\to\infty$: in the
Mott phase ($\varepsilon_{0}/U=0$) it tends to a finite value, whereas
it diverges at the Mott-BEC boundary (their position in the phase
diagram is shown in Figure \ref{fig:susceptibility}(b)). Note the scale
of the figure and how the susceptibility values are significantly
different already for the largest modest $N_{max}$ used ($=6)$.
Note that the dependence on $N_s$ is extremely weak, 
since it modifies only slightly the single-particle orbital in which
the bosons condense, whereas $N_{max}$ limits the maximum
number of bosons allowed to condense.

\begin{figure}[h!]
\includegraphics[scale=0.33]{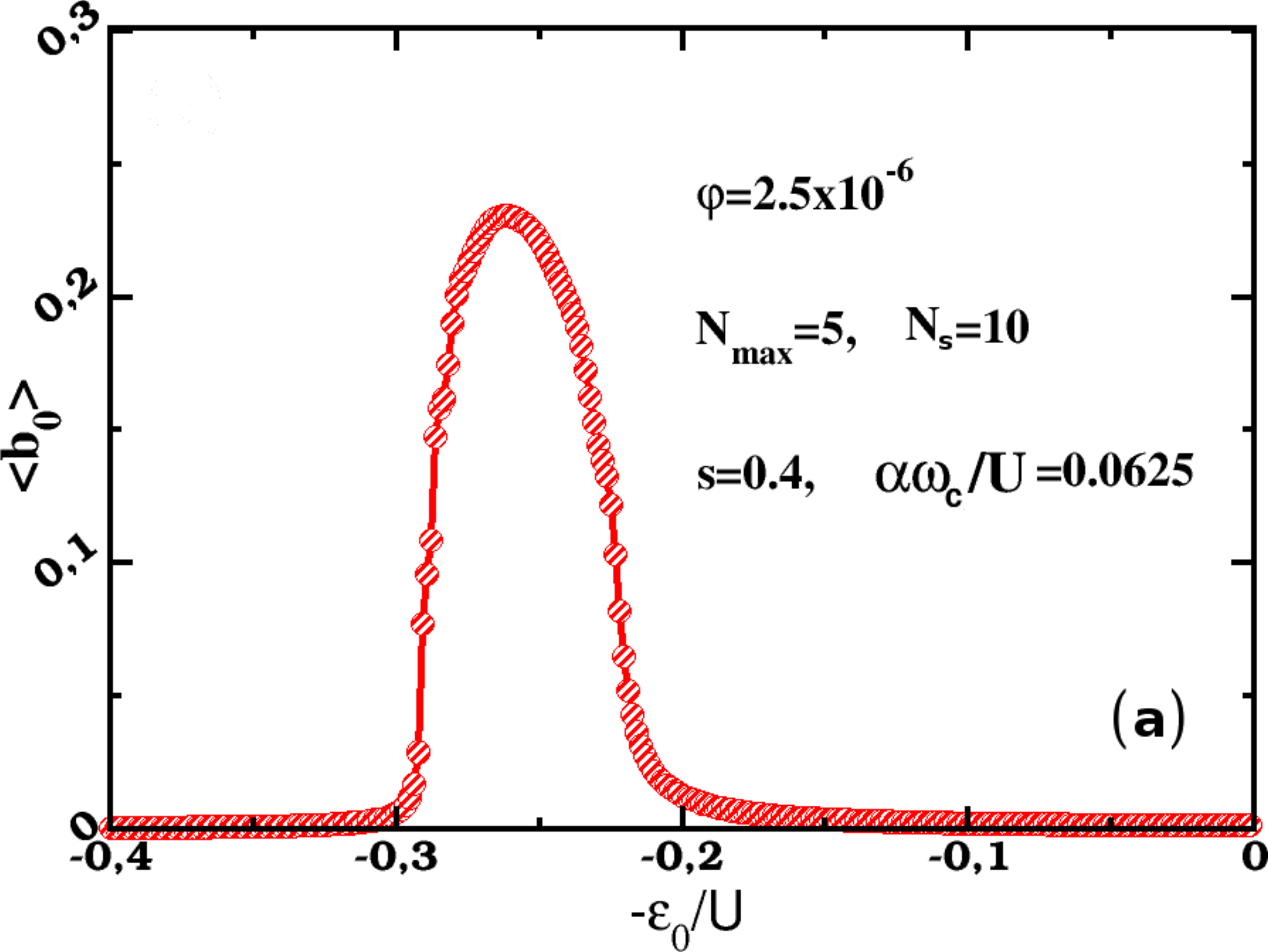}

\includegraphics[scale=0.33]{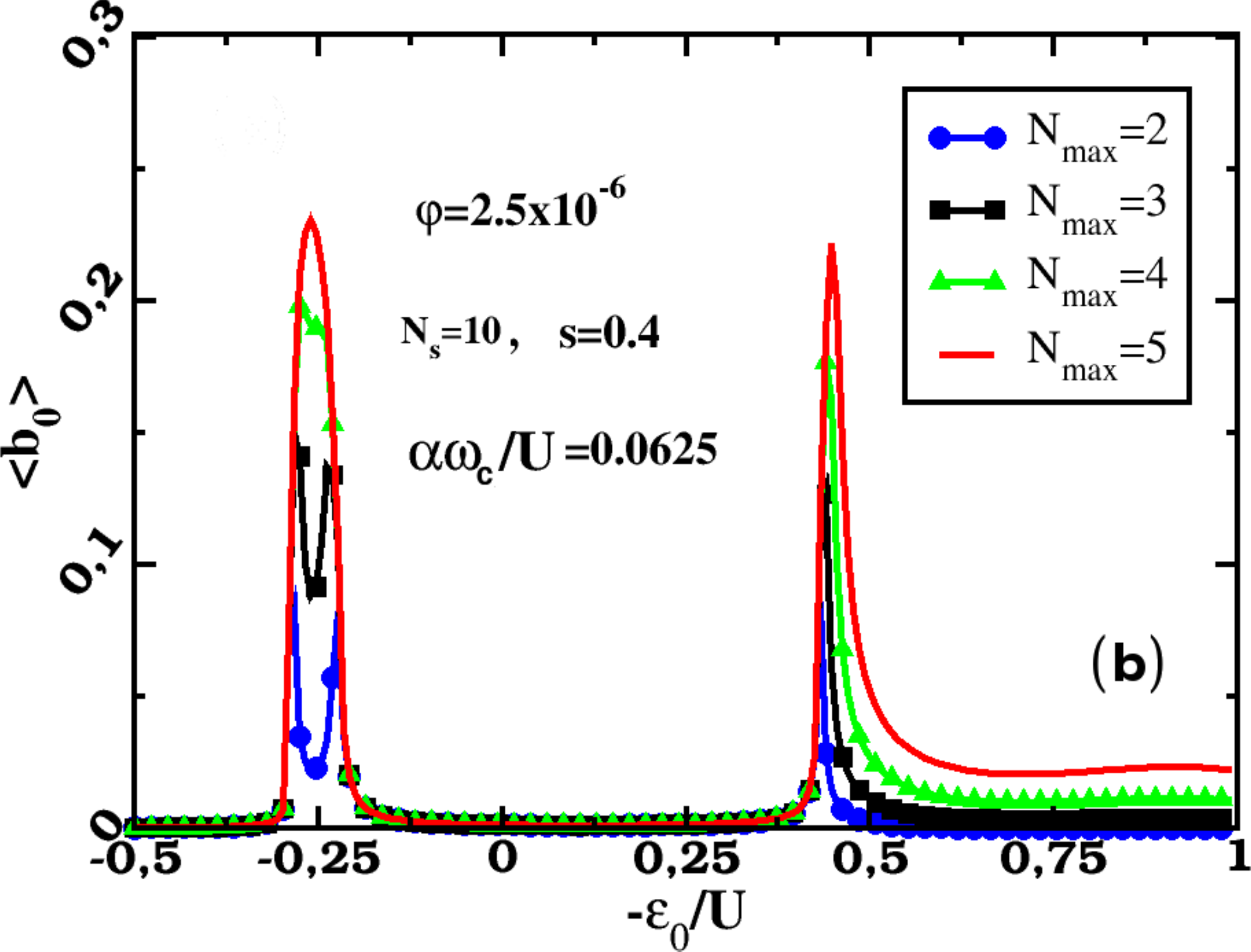}\caption{\label{fig:orderparameter} Behavior of the order parameter
$\left\langle b_{0}\right\rangle $ for a fixed small symmetry breaking
field ($\varphi\sim10^{-6}$) in the different phases. (a) $\left\langle b_{0}\right\rangle $
is finite inside the BEC phase but negligibly small within the Mott
phase; (b) the actual value of $\left\langle b_{0}\right\rangle $
is strongly dependent on the the truncation parameter $N_{max}$ within
the BEC phase. Other parameters are the same as in Figure \ref{fig:bversusphi}.}
\end{figure}

The great differences in susceptibility values allow us to determine
the phase diagram with fairly good accuracy also by fixing $\varphi$
at a small value, say $\sim10^{-6}$, and $N_{max}$ to a moderately
large, yet numerically feasible value of 5, and scanning $\varepsilon_{0}$
and $\alpha$. As an example of this procedure, we show in Figure \ref{fig:orderparameter}(a)
the order parameter as a function of $\varepsilon_{0}$ for
fixed $\alpha$ (this corresponds to the brown vertical line of Figure
\ref{fig:susceptibility}(b)). The BEC phase is clearly demarcated
from the Mott phase by a finite value of the order parameter. Note,
however, that the actual numerical value of $\left\langle b_{0}\right\rangle $
is strongly dependent on the truncation parameter $N_{max}$. Indeed,
this is clearly demonstrated in Figure \ref{fig:orderparameter}(b),
which shows a similar scan of $\varepsilon_{0}$ for different values
of $N_{max}$. For small values of $N_{max}$ the order parameter
shows pronounced peaks close to the phase boundaries but dips to smaller
values deep inside the BEC phase. Only for sufficiently large $N_{max}$
does one recover the single hump behavior of $\left\langle b_{0}\right\rangle $,
with a maximum value deep inside the BEC phase. Note that the criterion of a 
phase susceptibility of $10^4$ used in Figure \ref{fig:phasediagram} 
is equivalent to a value of the order parameter of $2.5 \times 10^{-2}$
in Figure \ref{fig:orderparameter}(a). 
This dependence on $N_{max}$ is not unexpected
since, as will be shown later, the fluctuations in the occupancy are
larger in the BEC phase. Thus, the truncation at a finite value of
the total number of bosons introduces important errors and correspondingly
larger values of $N_{max}$ are required for a good description in
this region. This allows us to estimate
the error in the determination of the phase boundary as 
$\Delta \varepsilon_{0c} \sim 0.05 U_0$.

Furthermore, for the value of $\alpha$ used in Figure
\ref{fig:orderparameter}, the system is always in a BEC phase for
$-\varepsilon_{0}/U\gtrsim0.3$. These large values of
$-\varepsilon_{0}$ give rise to large occupancies of the impurity
orbital. As a result, the description in this region is very poor
for the values of $N_{max}$ we employed, as can be seen in Figure \ref{fig:orderparameter}(b).

\subsection{Other observables}

\label{sub:other}

\begin{figure}[h!]
\includegraphics[bb=0bp 0bp 885bp 664bp,scale=0.4]{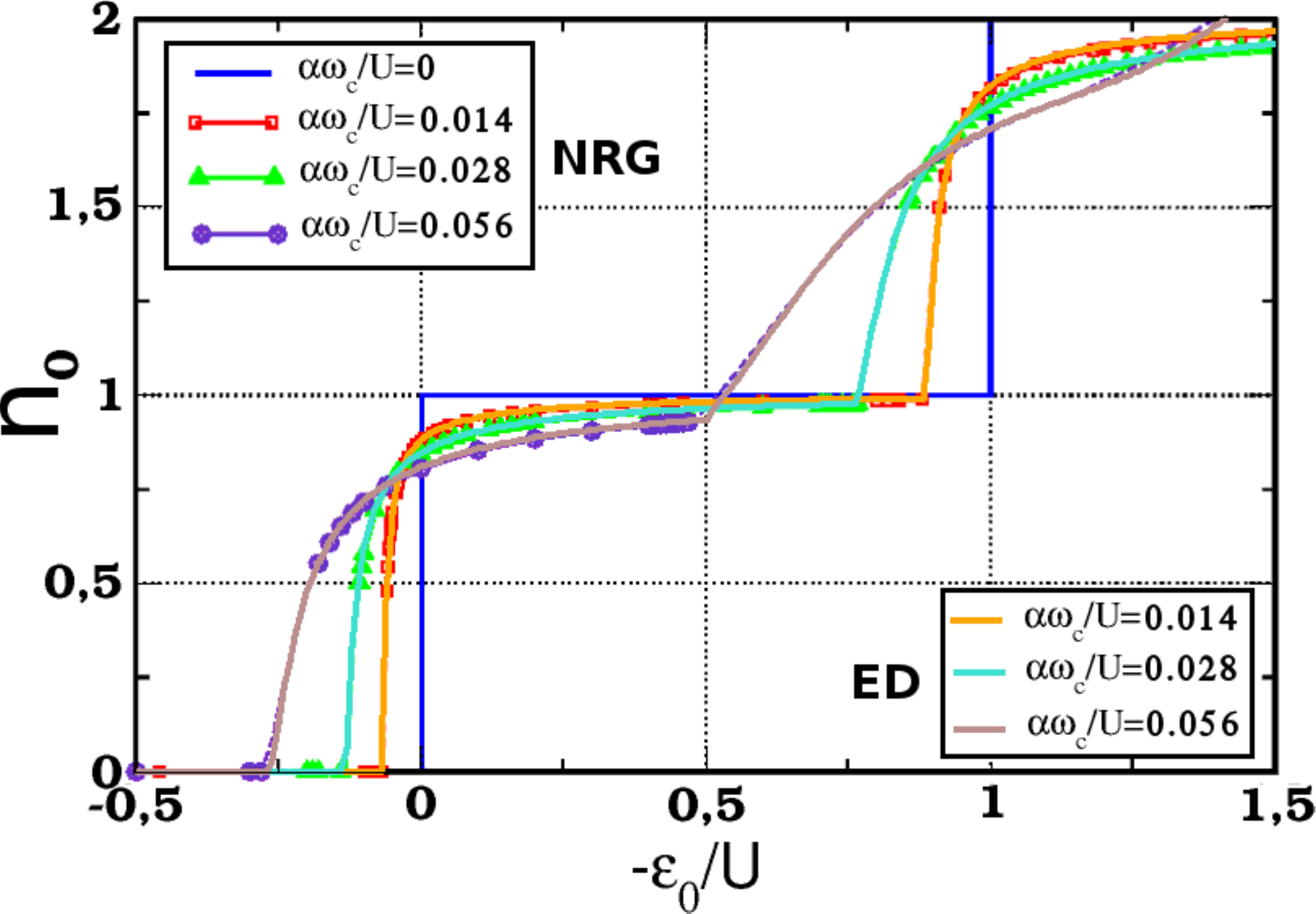} \caption{\label{fig:impoccupancy} Occupancy of the impurity
orbital as a function of $\varepsilon_{0}$ for different values of
$\alpha$. The symbols are the NRG results of reference \cite{Lee2007,Lee2010}
and the full lines correspond to our results. The agreement is remarkable.
The curves smoothly tend to the step-like behavior of the limit of
a decoupled impurity ($\alpha=0$). Other parameters are the same
as in Figure \ref{fig:bversusphi}.}
\end{figure}

As a test of the accuracy of our procedure, we have calculated other
local observables of the impurity orbital. Whenever available, we
have compared them with the NRG results \cite{Lee2007,Lee2010}. In
Figure \ref{fig:impoccupancy}, the impurity occupancy ($n_{0}=\langle b_{0}^{\dagger}b_{0}^{\phantom{\dagger}}\rangle$)
is shown for different values of the coupling to the bosonic bath
as a function of $\varepsilon_{0}$. Our results are the full lines
and the NRG results \cite{Lee2007} correspond to the symbols. The
regions without any symbol indicate the BEC phases. The agreement
is excellent and one can hardly distinguish the two sets of results.
As the coupling $\alpha$ to the bosonic bath decreases, the occupancy
tends smoothly to the step-like behavior of the decoupled impurity,
also shown Figure \ref{fig:impoccupancy} (blue line). It is this `adiabatic'
continuity between the Mott phases at $\alpha=0$ and $\alpha\neq0$
which allows us to ascribe a definite occupancy to the Mott `lobes'
of the phase diagram (see Figure \ref{fig:phasediagram}), even though
the occupancy is never exactly an integer for $\alpha>0$, as can
be seen in Figure \ref{fig:impoccupancy}. The excellent agreement shows
that our truncated Hilbert space calculation is more than enough for
a good description of at least some of the physical properties of
the impurity model.

\begin{figure}[h!]
\includegraphics[scale=0.35]{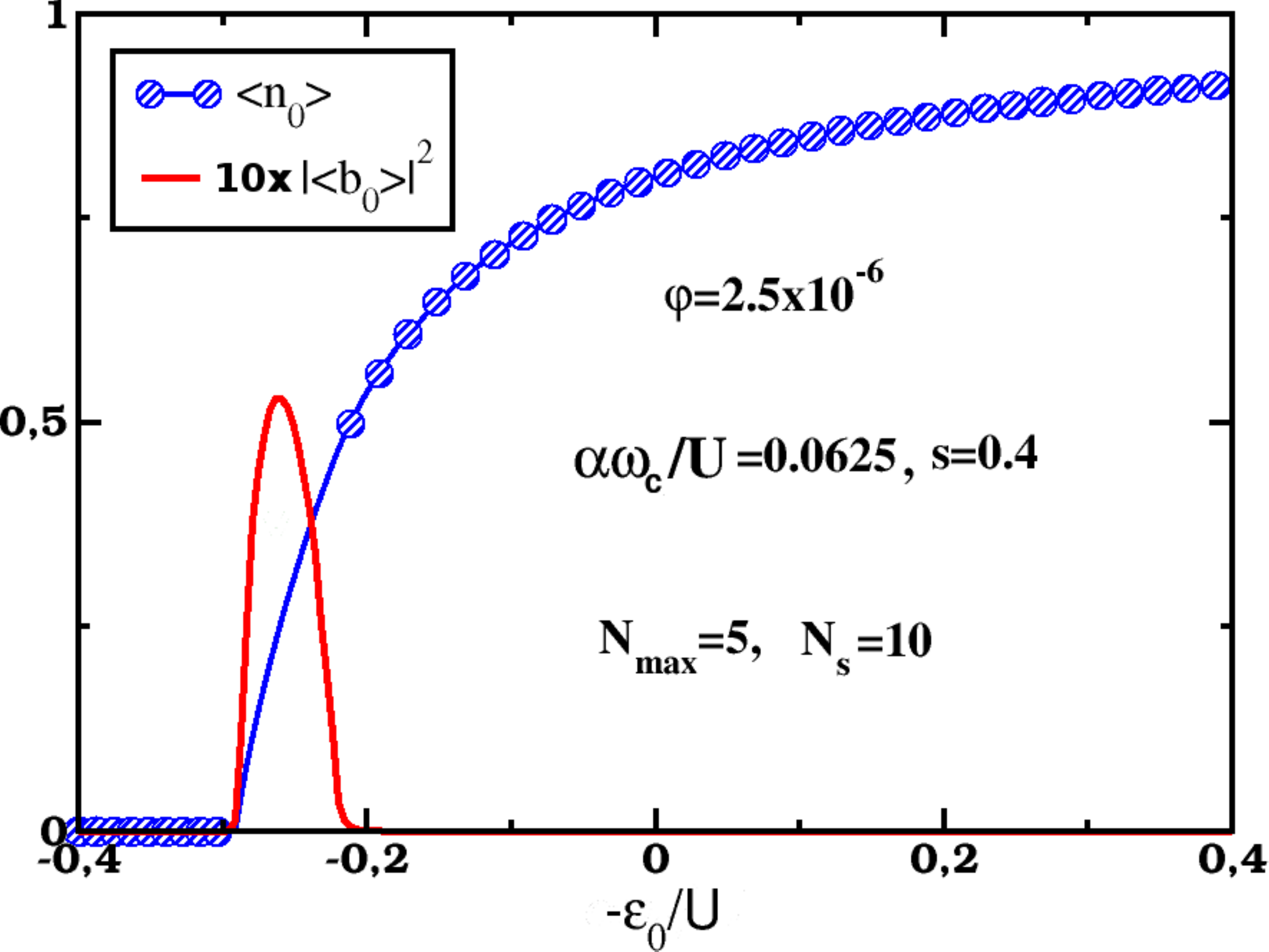} \caption{\label{fig:impoccversusb} Impurity occupancy and order
parameter squared as functions of the impurity single-particle energy $\varepsilon_{0}$.
The small symmetry breaking field hardly affects the occupancy. Other
parameters are the same as in Figure \ref{fig:bversusphi}.}
\end{figure}

For completeness, in Figure \ref{fig:impoccversusb} we show our results
for both the impurity occupancy (symbols) and the order parameter squared
(full line without symbols) as functions of $\varepsilon_{0}/U$,
both in the presence of a small symmetry-breaking field. We have rescaled
$\left|\left\langle b_{0}\right\rangle\right|^{2} $ by a factor of ten for greater
clarity. It is clear that the small symmetry-breaking field, although
essential to delineate the phases, affects very little the impurity
occupancy (compare with Figure \ref{fig:impoccupancy}). In this figure,
the absence of symbols in the BEC region is meant to mimic the convention
for this phase used in the NRG calculation \cite{Lee2007} (see Figure
\ref{fig:impoccupancy} for comparison).

Another striking feature highlighted in Figure \ref{fig:impoccversusb}
is the discrepant behavior of the impurity occupancy at the two borders
of the BEC phase. Whereas at the low $-\varepsilon_{0}$ border $n_{0}$
exhibits a discontinuity in its first derivative, the behavior at
the high $-\varepsilon_{0}$ border is perfectly smooth. This behavior
is generic to all the other BEC phases, as can be seen in Figure \ref{fig:impoccupancy}.
We conclude that although $n_{0}$ shows signs of critical behavior
at the low $-\varepsilon_{0}$ border of the BEC phases, it is not
critical at the other one.

\begin{figure}[h!]
\includegraphics[scale=0.4]{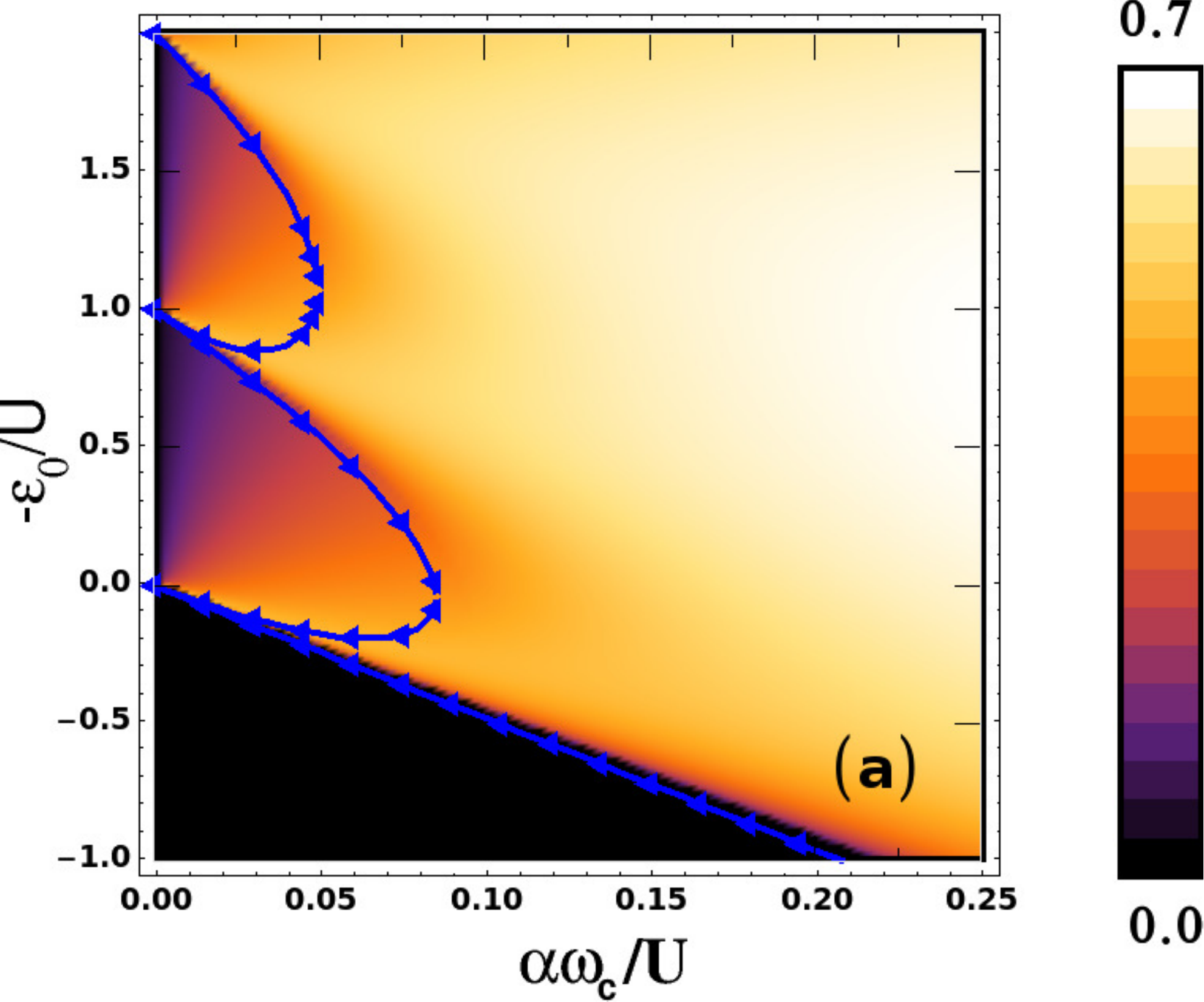}

\includegraphics[scale=0.34]{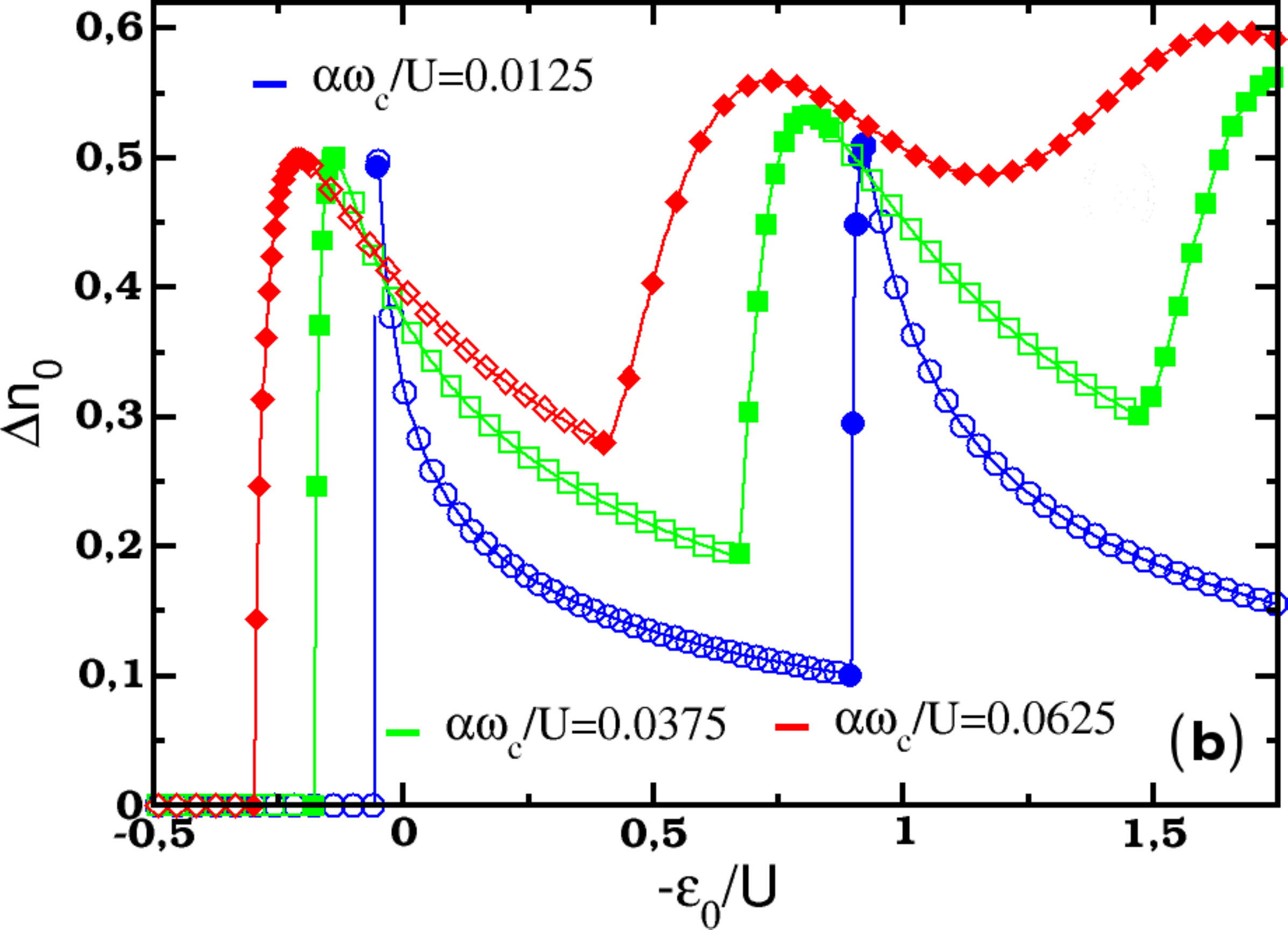}\caption{\label{fig:impoccfluct} Impurity occupancy fluctuation
$\Delta n_{0}$: (a) in the ($\varepsilon,\alpha$) plane. The full
blue line with symbols marks the phase boundaries); (b) as a function
of $\varepsilon_{0}$ for three values of $\alpha$. The open symbols
correspond to the `Mott' phase and closed ones to the BEC. Other parameters
are the same as in Figure \ref{fig:bversusphi}.}
\end{figure}

Finally, in Figure \ref{fig:impoccfluct}(a), we show the impurity occupancy
fluctuation $\Delta n_{0}=\sqrt{\left\langle \hat{n}_{0}^{2}\right\rangle -\left\langle \hat{n}_{0}\right\rangle ^{2}}$
in the ($\varepsilon_{0},\alpha$) plane with a color scale. The borders
between the phases are also shown as the blue lines with symbols.
Three vertical cuts across this plot are shown in Figure \ref{fig:impoccfluct}(b),
with the BEC region indicated by the closed symbols and the `Mott'
phase by the open ones. The occupancy fluctuation is generally larger
in the BEC regions as compared to the `Mott' phases, as expected.
Indeed, this is compatible with the requirement of larger values of
$N_{max}$ for a better description deep within the BEC phases (see
Figure \ref{fig:orderparameter}b).

However, a more thorough inspection shows that whereas $\Delta n_{0}$
increases rapidly upon entering the BEC phase through its low $-\varepsilon_{0}$
border, it goes through a maximum while still \emph{inside} the BEC
phase. Upon further increasing $-\varepsilon_{0}$, $\Delta n_{0}$
decreases until it finally crosses the high $-\varepsilon_{0}$ border
of the BEC phase, where it shows no sign of critical behavior, in
close similarity with the behavior of $n_{0}$ shown in Figure \ref{fig:impoccversusb}.
We stress that, even though the high $-\varepsilon_{0}$ transition
does not manifest itself in these quantities, does not mean that 
the system does not experience a true phase transition in that region.
Furthermore, the occupancy fluctuation does not vanish and is always
a monotonically decreasing function of $-\varepsilon_{0}$ in the
`Mott' phases for $n_{0}\neq0$.

\section{Discussion and conclusions}

\label{sec:Conclusions}

In this article, we have fully characterized the diverse physical
properties of the B-SIAM by using the relatively undemanding method
of exact diagonalization of small clusters. Besides, we have shown
how the physically motivated criterion of a spontaneously broken gauge
symmetry can be used to accurately identify the BEC or `Mott' phases
of the B-SIAM, even with a fairly small truncated Hilbert space. This
serves as a proof of principle of the criterion in this particular
case. Clearly, the detailed quantum critical behavior requires the
use of many more states, in which case the NRG method is probably
indispensable. However, the exact diagonalization method is accurate
enough for the determination of phase diagrams and physical properties,
as we have shown. This is particularly important for applications
such as the disordered version of BDMFT. In these cases, the use
of the more accurate NRG method is prohibitive.

We have also uncovered an unnoticed asymmetry in the critical behavior
of local quantities as one goes from `Mott' to BEC and BEC to `Mott'.
In the former case, both the mean occupancy and its fluctuations exhibit
a discontinuity in the first derivative with respect to the impurity
energy, whereas the latter transition seems to be completely smooth.
This is
probably a feature of the single-impurity model only, however, since
the incompressible nature of the Mott phase in the lattice case requires
both borders to show non-analytic behavior. The single-impurity model,
on the other hand, is not incompressible in the `Mott' phase.

\section*{Acknowledgments}

This work was supported by CNPq through grants 304311/2010-3 (EM)
and 140184/2007-4 (JHW) and by FAPESP through grant 07/57630-5 (EM). 

\appendix

\renewcommand{\theequation}{A\arabic{equation}}
\setcounter{equation}{0}

\section{General properties of the non-interacting limit}

\label{sec:non-interacting}

\subsection{Diagonalization of the non-interacting model}

We will first briefly describe the diagonalization of the non-interacting
Hamiltonian, Equation \eqref{eq:hamiltonian} with $U$ set to zero. The
absence of interactions makes this process independent of the statistics,
since it amounts to finding the basis of single-particle states that
makes the Hamiltonian diagonal. A more detailed calculation, in the
language of fermions, can be found in reference \cite{Mahan4.2}.
We define annihilation operators $c_{j}$ for the diagonal single-particle
states through

\begin{eqnarray}
b_{0}= & \sum_{j} & \gamma_{j}c_{j},\label{eq:gammaj}\\
b_{k}= & \sum_{j} & \nu_{kj}c_{j},\label{eq:nukj}
\end{eqnarray}
so that the non-interacting Hamiltonian can be written in this basis
as 
\begin{equation}
H=\sum_{j}E_{j}c_{j}^{\dagger}c_{j}^{\phantom{\dagger}}.\label{eq:diagonalham}
\end{equation}
The coefficients $\gamma_{j}$ and $\nu_{kj}$ have yet to be determined.
By taking the commutator of $b_{k}$ and $c_{j}$ with the Hamiltonian
in the forms \eqref{eq:hamiltonian} and \eqref{eq:diagonalham-2},
respectively, and using \eqref{eq:gammaj} and \eqref{eq:nukj}, we
find the eigenvalue system of equations for the unknown coefficients

\begin{eqnarray}
\left(E_{j}-\varepsilon_{0}\right)\gamma_{j} & = & \sum_{k}V_{k}\nu_{kj},\label{eq:gammajeq}\\
\left(E_{j}-\varepsilon_{k}\right)\nu_{kj} & = & V_{k}\gamma_{j}.\label{eq:nukjeq}
\end{eqnarray}
 Inserting Equation \eqref{eq:nukjeq} into Equation \eqref{eq:gammajeq} we
obtain an implicit equation for the eigenvalues $E_{j}$
\begin{equation}
E_{j}-\varepsilon_{0}=\sum_{k}\frac{V_{k}^{2}}{E_{j}-\varepsilon_{k}}.\label{eq:eigenvalues}
\end{equation}
The normalization condition
\begin{equation}
\gamma_{j}^{2}+\sum_{k}\nu_{kj}^{2}=1,\label{eq:normalization}
\end{equation}
 together with \eqref{eq:nukjeq}, leads to
\begin{equation}
\gamma_{j}^{-2}=1+\sum_{k}\dfrac{V_{k}^{2}}{\bigl(E_{j}-\varepsilon_{k}\bigr)^{2}},\label{eq:gammajsol}
\end{equation}
and finally Equation \eqref{eq:nukjeq} can be used to find an expression
for the $\nu_{kj}$ coefficients, which we will omit.

\subsection{Critical coupling for a power-law spectral density}

The BEC in the non-interacting model occurs when the lowest eigenvalue
(measured with respect to the chemical potential) vanishes. From Equation
\eqref{eq:eigenvalues}, this happens when
\begin{equation}
\varepsilon_{0}=\sum_{k}\frac{V_{k}^{2}}{\varepsilon_{k}}.\label{eq:criticalequation}
\end{equation}
We would like to analyze this equation in the continuum limit. In
this case, we can use the function defined in Equation \eqref{eq:bathfunction},
which satisfies the Kramers-Kronig relation
\begin{equation}
\mathrm{Re}\Delta\left(\omega-i\delta\right)=\fint\frac{dx}{\pi}\frac{\mathrm{Im}\Delta\left(x-i\delta\right)}{\omega-x},\label{eq:Kramers-Kronig}
\end{equation}
in which the integral is a principal part. Thus, one may replace the
right-hand side of Equation \eqref{eq:criticalequation} by
\[
\sum_{k}\frac{V_{k}^{2}}{\varepsilon_{k}}=-\mathrm{Re}\Delta\left(0-i\delta\right)=\fint\frac{dx}{\pi}\frac{\mathrm{Im}\Delta\left(x-i\delta\right)}{x}.
\]
For the power-law spectral function of Equation \eqref{eq:spectralfunction},
we are left with an expression for the critical coupling $\alpha_{c}^{0}$
\[
\varepsilon_{0}=2\alpha_{c}^{0}\omega_{c}^{1-s}\int_{0}^{\omega_{c}}x^{s-1}dx=\frac{2}{s}\alpha_{c}^{0}\omega_{c}\ \ \left(s>0\right).
\]
Note that the equation is not well defined for $s\leq0$. We finally
find that
\[
\alpha_{c}^{0}=\frac{s\varepsilon_{0}}{2\omega_{c}}\ \ \left(s>0\right).
\]

\subsection{The non-interacting Hamiltonian in the presence of a symmetry-breaking
field}

We now consider the non-interacting Hamiltonian in the presence of
a symmetry-breaking field, Equation \eqref{eq:hamiltonianphi}. We first
define `displaced' operators $a_{0}$ and $a_{k}$
\begin{eqnarray}
b_{0} & = & a_{0}+\lambda,\label{eq:a0}\\
b_{k} & = & a_{k}+\eta_{k},\label{eq:ak}
\end{eqnarray}
where the parameters $\lambda,\eta_{k}$ can be taken to be real without
loss of generality. Inserting these into \eqref{eq:hamiltonianphi}
we end up with
\begin{eqnarray*}
H & = & \varepsilon_{0}a_{0}^{\dagger}a_{0}^{\phantom{\dagger}}+\left(\varphi+\lambda\varepsilon_{0}+\sum_{k}\eta_{k}V_{k}\right)\left(a_{0}^{\dagger}+a_{0}^{\phantom{\dagger}}\right)\\
 & + & \sum_{k}\varepsilon_{k}a_{k}^{\dagger}a_{k}^{\phantom{\dagger}}+\sum_{k}\left(\varepsilon_{k}\eta_{k}+\lambda V_{k}\right)\left(a_{k}^{\dagger}+a_{k}^{\phantom{\dagger}}\right)\\
 & + & \sum_{k}V_{k}\left(a_{k}^{\dagger}a_{0}^{\phantom{\dagger}}+a_{0}^{\dagger}a_{k}^{\phantom{\dagger}}\right)\\
 & + & \varepsilon_{0}\lambda^{2}+2\lambda\varphi+\sum_{k}\varepsilon_{k}\eta_{k}^{2}+2\lambda\sum_{k}V_{k}\eta_{k}.
\end{eqnarray*}
 The terms linear in the new `displaced' operators can be eliminated
if we choose $\lambda$ and $\eta_{k}$ to satisfy
\begin{eqnarray}
\varphi+\lambda\varepsilon_{0}+\sum_{k}\eta_{k}V_{k} & = & 0,\label{eq:dispeq1}\\
\varepsilon_{k}\eta_{k}+\lambda V_{k} & = & 0.\label{eq:dispeq2}
\end{eqnarray}
 Taking $\eta_{k}$ from Equation \eqref{eq:dispeq2} into \eqref{eq:dispeq1}
we find
\begin{eqnarray}
\eta_{k} & = & -\lambda\dfrac{V_{k}}{\varepsilon_{k}},\label{eq:etak}\\
\lambda & = & \dfrac{\varphi}{\sum_{k}\frac{V_{k}^{2}}{\varepsilon_{k}}-\varepsilon_{0}}\equiv\frac{\varphi}{\kappa}.\label{eq:lambda}
\end{eqnarray}
For this choice, we are left with 
\begin{equation}
H=\varepsilon_{0}a_{0}^{\dagger}a_{0}^{\phantom{\dagger}}+\sum_{k}\varepsilon_{k}a^{\dagger}a_{k}^{\phantom{\dagger}}+\sum_{k}V_{k}(a_{k}^{\dagger}a_{0}^{\phantom{\dagger}}+a_{0}^{\dagger}a_{k}^{\phantom{\dagger}})+\frac{\varphi^{2}}{\kappa}.\label{eq:displacedham}
\end{equation}

As shown before, the Hamiltonian \eqref{eq:displacedham} can be brought
to diagonal form by means of the canonical transformation of Eqs.
\eqref{eq:gammaj} and \eqref{eq:nukj}, with $\gamma_{j}$ given
by \eqref{eq:gammajsol} and $\nu_{kj}$ given by \eqref{eq:nukjeq}.
Thus,
\begin{equation}
H=\sum_{j}E_{j}c_{j}^{\dagger}c_{j}^{\phantom{\dagger}}+\frac{\varphi^{2}}{\kappa}.\label{eq:diagonalham-2}
\end{equation}
where,
\begin{eqnarray}
b_{0} & = & \lambda+\sum_{j}\gamma_{j}c_{j},\label{eq:b0toc}\\
b_{k} & = & \eta_{k}+\sum_{j}\nu_{kj}c_{j}.\label{eq:bktoc}
\end{eqnarray}

Since the ground state $\vert\Phi_{0}\rangle$ is annihilated by the
$c_{j}$ operators it follows that
\begin{eqnarray}
\sum_{j}\gamma_{j}c_{j}\vert\Phi_{0}\rangle & = & (b_{0}-\lambda)\vert\Phi_{0}\rangle=0,\label{eq:ckillsgs-1}\\
\sum_{j}\nu_{kj}c_{j}\vert\Phi_{0}\rangle & = & (b_{k}-\eta_{k})\vert\Phi_{0}\rangle=0.\label{eq:ckillsgs-2}
\end{eqnarray}
Therefore, the expectation values of the $b$ operators in the ground
state does not vanish
\begin{eqnarray}
\langle b_{0}\rangle & = & \lambda,\label{eq:b0exp}\\
\langle b_{k}\rangle & = & \eta_{k.}\label{eq:bkexp}
\end{eqnarray}
Furthermore, the total number of particles is given by
\begin{equation}
N=\langle b_{0}^{\dagger}b_{0}^{\phantom{\dagger}}\rangle+\sum_{k}\langle b_{k}^{\dagger}b_{k}^{\phantom{\dagger}}\rangle=\lambda^{2}+\sum_{k}\eta_{k}^{2}.\label{eq:ntotal}
\end{equation}
Using Equations \eqref{eq:etak} and \eqref{eq:lambda}
\begin{equation}
N=\left(1+\sum_{k}\frac{V_{k}^{2}}{\varepsilon_{k}^{2}}\right)\frac{\varphi^{2}}{\kappa^{2}}.\label{eq:ntotal-2}
\end{equation}
%
%
%



\end{document}